\documentclass[10pt,letterpaper]{article}
\usepackage{opex3}
\usepackage{amsmath}            
\usepackage{amssymb}            
\usepackage{amscd}              
\usepackage{dsfont}              
\usepackage{amsbsy}

\begin{document}

\title{Threshold of a random laser based on Raman gain in cold atoms}

\author{William Guerin, Nicolas Mercadier, Davide Brivio$^1$ and Robin Kaiser}

\address{Institut Non Lin\'{e}aire de Nice, CNRS and Universit\'{e} de
Nice Sophia-Antipolis,\\ 1361 route des Lucioles, 06560 Valbonne,
France}
\address{$^1$Present address: Dipartimento di Fisica, Universit\`a di Milano, I-20113, Italy}
\email{william.guerin@inln.cnrs.fr} 
\homepage{http://www.kaiserlux.de/coldatoms/} 


\begin{abstract}
We address the problem of achieving a random laser with a cloud of
cold atoms, in which gain and scattering are provided by the same
atoms. In this system, the elastic scattering cross-section is
related to the complex atomic polarizability. As a consequence,
the random laser threshold is expressed as a function of this polarizability,
which can be fully determined by spectroscopic measurements. We
apply this idea to experimentally evaluate the threshold of a
random laser based on Raman gain between non-degenerate Zeeman
states and find a critical optical thickness on the order of 200,
which is within reach of state-of-the-art cold-atom experiments.
\end{abstract}

\ocis{(140.1340) Atomic gas lasers; (140.3550) Lasers, Raman; (290.4210) Multiple scattering} 




\section{Introduction}

Random lasing occurs when the optical feedback due to multiple
scattering (or ``radiation trapping") in a gain medium is strong
enough so that gain in the sample volume overcomes losses through
the surface. Since its theoretical prediction by Letokhov
\cite{Letokhov:1968}, great efforts have been made to
experimentally demonstrate this effect in different kinds of
systems
\cite{Gouedard:1993,Martorell:1996,Cao:1998,Wiersma:2001,Strangi:2006},
as well as to understand the fundamentals of random lasing
\cite{Wiersma:1996,Burin:2001,Vanneste:2007,Tureci:2008}. The broad interest of
this topic is driven by potential applications (see
\cite{Wiersma:2008} and references therein) and by its connections
to the fascinating subject of Anderson localization
\cite{Conti:2008}. State-of-the-art random lasers
\cite{Wiersma:2008,Cao:2003,Cao:2005} are usually based on
condensed matter systems, and feedback is provided by a disordered
scattering medium, while gain is provided by an active material
lying in the host medium or inside the scatterers. In general,
scattering and gain are related to different physical entities.

Another system that can be considered for achieving random lasing
is a cold atomic vapor, using magneto-optical traps
\cite{Metcalf}, where radiation trapping
\cite{Fioretti:1998,Labeyrie:2003} as well as lasing
\cite{Hilico:1992,Guerin:2008} have already been demonstrated. One
advantage is the ability to characterize and model the microscopic
properties of the medium, which can be extremely valuable for a
better understanding of the physics of random lasers.

However, in such system, the ability to combine gain and multiple
scattering at the same time is not obvious, as both should be
provided by the same atoms. On the other hand, it has been shown
recently that the peculiarity of this system leads to a simple
condition for random lasing in the incoherent regime
\cite{footnote1}. The threshold is indeed defined as a critical
on-resonance optical thickness $b_0$, which is a function of the
complex atomic polarizability $\alpha$ as the single parameter
\cite{Froufe:2009}. This has been used to predict theoretically
the threshold of a random laser based on Mollow gain, for which
the atomic polarizability is analytically known
\cite{Froufe:2009,Mollow:1972}. A critical $b_0$ of the order of
300 has been found.

In contrast to the {\it ab initio} theoretical approach of
\cite{Froufe:2009}, we present here an experimental evaluation of
the threshold of a random laser. Our method relies on the fact
that thanks to Kramers-Kronig relations \cite{Jackson}, the
complex atomic polarizability is indeed one \emph{single}
independent parameter, and thus can be fully determined by a
spectroscopic measurement. This idea is general and could be
applied with any gain mechanisms. We demonstrate its usefulness
here with Raman gain between non-degenerate Zeeman states
\cite{Hilico:1992,Guerin:2008,Grison:1991,Tabosa:1991}. We obtain
a critical optical thickness on the order of 200, lower than with
Mollow gain \cite{Froufe:2009}.

\section{Measuring the threshold of a random laser with cold atoms}

From Letokhov's diffusive description of light transport in a
homogeneous, disordered and active medium of size $L$, we know
that the random laser threshold is governed by two characteristic
lengths: the elastic scattering mean free path $\ell_\mathrm{sc}$
\cite{Rossum:1999,footnote2} and the linear gain length
$\ell_\mathrm{g}$ ($\ell_\mathrm{g}<0$ corresponds to absorption
or inelastic scattering). In the diffusive regime, defined as $L
\gg \ell_\mathrm{sc}$, the lasing threshold is reached when the
unfolded path length, on the order of $L^2/\ell_\mathrm{sc}$,
becomes larger than the gain length. More precisely, the threshold
is given by \cite{Letokhov:1968,Cao:2003}
\begin{equation}\label{eq.letokhov}
L_\mathrm{eff} > \beta \pi \sqrt{\ell_\mathrm{sc}\,
\ell_\mathrm{g} /3} \; ,
\end{equation}
where $\beta$ is a numerical factor that depends on the geometry
of the sample ($\beta=1$ for a slab, $\beta=2$ for a sphere, which
is the case we consider in the following), and $L_\mathrm{eff} =
\eta L$ is the effective length of the sample, taking into account
the extrapolation length \cite{Rossum:1999}. For $L >
\ell_\mathrm{sc}$ and a sphere geometry, $\eta = 1 + 2 \xi /
\left[L/\ell_\mathrm{sc} + 2\xi \right]$ with $\xi \simeq 0.71$
\cite{zweifel,Drozdowicz:2003}. Note that deeply in the diffusive
regime ($L \gg \ell_\mathrm{sc}$), $\eta \sim 1$. Another
important length scale is the extinction length
$\ell_\mathrm{ex}$, as measured by the forward transmission of a
beam through the sample, $T = e^{-L/\ell_\mathrm{ex}}$. The
extinction length is related to the other lengths by
$\ell_\mathrm{ex}^{-1} = \ell_ \mathrm{sc}^{-1}
-\ell_\mathrm{g}^{-1}$.

For an atomic vapor, these characteristic lengths can both be
computed as a function of the atomic polarizability
$\alpha(\omega)$ at frequency $\omega$. The extinction
cross-section is indeed given by $\sigma_\mathrm{ex}(\omega) = k
\times \mathrm{Im}[\alpha(\omega)]$ and the elastic scattering
cross-section by $\sigma_\mathrm{sc}(\omega) = k^4/6\pi \times
|\alpha(\omega)|^2$ \cite{Lagendijk:1996} ($k=\omega/c$ is the
wave vector). Note that the first relation is general to any
dielectric medium whereas the second one is specific to resonant
point-dipole scatterers. The characteristic lengths are then
$\ell_\mathrm{ex,sc}^{-1} = \rho\, \sigma_\mathrm{ex,sc}$, where
$\rho$ is the atomic density. The gain cross-section can be
defined the same way by $\ell_\mathrm{g}^{-1} =
\rho\,\sigma_\mathrm{g}$. The vapor is supposed homogeneous, as
well as the pumping field, so that both $\rho$ and $\alpha$ are
position-independent. Even though this is not the precise geometry
of a cold-atom experiment, it allows us to perform analytical
estimations. As we consider only quasi-resonant light, we shall
use $k=k_0=\omega_0/c$ with $\omega_0$ the atomic eigenfrequency.
In the following, we shall also use a dimensionless atomic
polarizability $\tilde{\alpha}$, defined as $\alpha =
\tilde{\alpha} \times 6\pi/k_0^3$, and omit the dependence on
$\omega$. We can now rewrite $\sigma_\mathrm{sc} = \sigma_0
|\tilde{\alpha}|^2$ and $\sigma_\mathrm{g}=\sigma_0
\left(|\tilde{\alpha}|^2-\mathrm{Im}(\tilde{\alpha})\right)$,
where $\sigma_0 = 6\pi/k_0^2$ is the resonant scattering
cross-section, such that the threshold condition, as expressed by
Eq. (\ref{eq.letokhov}), reduces to
\begin{equation}\label{eq.b0cr} \rho \sigma_0 L_\mathrm{eff} =
\eta b_0  > \frac{2\pi}{\sqrt{3 |\tilde{\alpha}|^2 \, \left(
|\tilde{\alpha}|^2-\mathrm{Im}(\tilde{\alpha}) \right)}} \, ,
\end{equation} where $b_0$ is the on-resonance optical thickness
of the cloud.  This condition is valid as soon as the medium
exhibits gain, \textit{i.e.}
$|\tilde{\alpha}|^2-\mathrm{Im}(\tilde{\alpha}) > 0$.

The threshold condition is thus given by a critical on-resonance
optical thickness, which is an intrinsic parameter of the cloud,
expressed as a function of the complex atomic polarizability only,
which depends on the pumping parameters. Although the initial
condition of Eq. (\ref{eq.letokhov}) involves two characteristic
lengths, we emphasize here that this is really one single
independent parameter, as real and imaginary parts of the atomic
polarizability are related via Kramers-Kronig relations
\cite{Jackson}. This point is due to the originality of the system
that we are considering, in which the same atoms are used to
amplify and scatter light.

This property has two important practical consequences. The first
one is that we cannot adjust one quantity (for example gain)
independently of the other (scattering rate, or vice versa), so
that the existence of reasonable conditions for random lasing is
not obvious. This issue has been positively answered recently and
it has been shown that random lasing can even occur with a low
amount of scattering \cite{Froufe:2009}. The second one is that
only one quantity has to be measured to determine the threshold,
as soon as we can measure it for every $\omega$, since
Kramers-Kronig relations involve integrals over $\omega$. A weak
probe transmission spectrum, which we can rewrite, with our
notations, $T(\omega)= \exp\left[-b_0 \times
\mathrm{Im}\left(\tilde{\alpha}(\omega)\right)\right]$, contains
therefore enough information to fully characterize
$\tilde{\alpha}(\omega)$ and then to deduce the critical optical
thickness. In the following, we use this idea with Raman gain.

Note that without this possibility, measuring independently the
two characteristic lengths is difficult. Besides the transmission
spectrum, one needs another measurement, which can be provided by
the fluorescence. Nevertheless, the probe fluorescence is small
compared to the pump one, and inelastic scattering is not easily
distinguished from elastic scattering. Despite these difficulties,
preliminary measurements, in a limited range of parameters, have
qualitatively validated the approached based on Kramers-Kronig
relations \cite{Davide:Tesi}.

\section{Application to Raman gain}

\begin{figure}[t]
\centering
\includegraphics{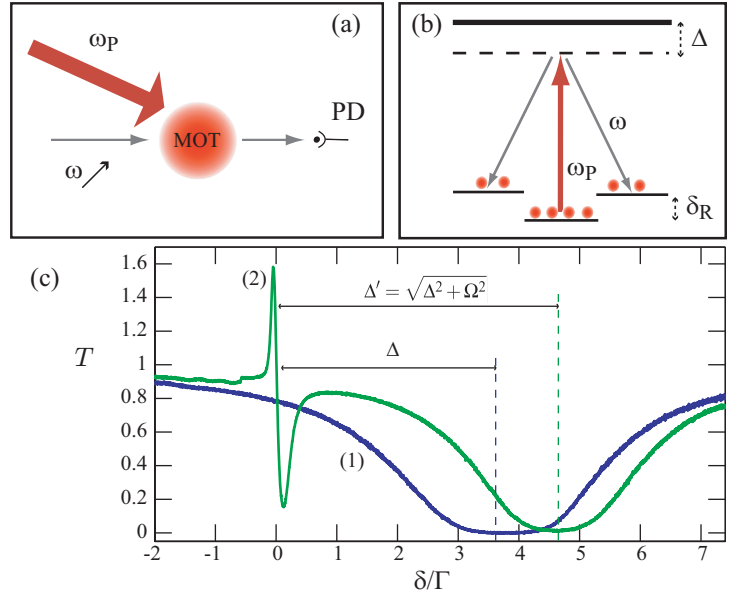}
\caption{(a) Principle of the experiment. We send a weak probe
beam on the magneto-optical trap (MOT) and the transmission is
recorded on a photodetector (PD). The probe frequency $\omega$ is
ramped during the acquisition in order to record a spectrum.
Another, stronger beam of frequency $\omega_\mathrm{P}$ is used as
a pump. (b) Principle of the Raman mechanism (depicted here for a
$F=1 \rightarrow F'=2$ transition). (c) Experimental transmission
spectra, plotted as a function of the pump-probe detuning
$\delta$. Without pumping, spectrum (1) shows only the atomic
absorption. A pump beam of detuning $\Delta=-3.8 \Gamma$ and
intensity $13$~mW/cm$^2$, corresponding to a Rabi frequency
$\Omega = 2.5 \Gamma$, is added to obtain spectrum (2), which then
exhibits a Raman resonance in the vicinity of $\delta=0$. The
atomic absorption is shifted due to the pump-induced light shift
and the absorption is reduced due to saturation.}\label{fig.Raman}
\end{figure}

Our experiment uses a cloud of cold $^{85}$Rb atoms confined in a
vapor-loaded magneto-optical trap (MOT) \cite{Metcalf} produced by
six large independent trapping beams, allowing the trapping of a
few $10^9$ atoms at a density of $10^{10}$~atoms/cm$^3$,
corresponding to an on-resonance optical thickness of about 10. To
add gain to our system, we use a pump beam, which is tuned near
the $F=3\rightarrow F'=4$ cycling transition of the $D2$ line of
$^{85}$Rb (frequency $\omega_0$, wavelength $\lambda = 780$~nm,
natural linewidth $\Gamma/2\pi = 6.1$~MHz), with a detuning
$\Delta=\omega_\mathrm{P}-\omega_0$, which can be changed via an
acousto-optic modulator in a double-pass configuration. The pump
beam has a linear polarization and a waist larger than the MOT
size (a few millimeters) to ensures homogeneous pumping. An
additional, orthogonally polarized beam is used as a weak probe to
measure transmission spectra with a propagation axis making an
angle with the pump-beam axis of about $17^\circ$ [Fig.
\ref{fig.Raman}(a)]. This small angle, together with the low
temperature of our sample ($\sim 100~\mu$K) allows us to neglect
any relative Doppler broadening ($\sim 40$~kHz). The probe
frequency $\omega$ can be swept around the pump frequency with a
detuning $\delta=\omega-\omega_\mathrm{P}$. Both lasers, pump and
probe, are obtained by injection-locking of semiconductor lasers
from a common master laser, which allows to resolve narrow
spectral features (this has been checked for earlier experiments
\cite{Guerin:2008} down to 10~kHz). All our experiments are
time-pulsed with a cycling time of 30~ms. The trapping period
lasts 29~ms, followed by a dark period of 1~ms, when the MOT
trapping beams and magnetic field are switched off. In order to
avoid optical pumping into the dark hyperfine $F=2$ ground state,
a repumping laser is kept on all time. Pump-probe spectroscopy is
performed during the dark phase, short enough to avoid expansion
of the atomic cloud. Data acquisitions are the result of an
average of 300 cycles.

\begin{figure}[t]
\centering
\includegraphics{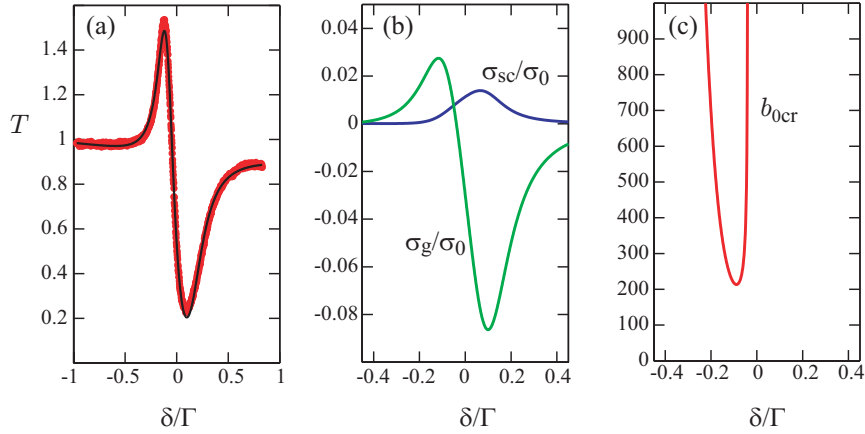} \caption{(a) Typical experimental spectrum (red dots) and its fit (black line) around the Raman
resonance.The parameters obtained from the fit are $A_1 = 0.21$
(gain amplitude), $A_2 = 0.11$ (absorption amplitude), $\gamma =
0.25 \Gamma = 1.5$~MHz and $\delta_\mathrm{R} = 0.09 \Gamma =
540$~kHz. (b) Gain and scattering cross sections, computed from
Eqs. (\ref{eq.sigma_g},\ref{eq.sigma_sc}) with the Raman
parameters deduced from the fit. (c) Corresponding critical
optical thickness. The minimum is $b_0 \simeq 220$. This set of
data corresponds to the pump parameters $\Delta=-3.4 \Gamma$ and
$\Omega = 3.4 \Gamma$. }\label{fig.fit}
\end{figure}

Raman gain relies on the pump-induced population inversion among
the different light-shifted $m_F$ Zeeman sublevels of the $F=3$
hyperfine level \cite{Grison:1991,Tabosa:1991}, as depicted in
Fig. \ref{fig.Raman}(b). The optical pumping induced by the
$\pi$-polarized pump laser leads to a symmetric distribution of
population with respect to the $m_F=0$ sublevel of the ground
state, with this sublevel being the most populated and also the
most shifted, due to a larger Clebsch-Gordan coefficient
\cite{Brzozowski:2005}. Atoms are probed with a $\pi$-polarized
(with perpendicular direction) probe beam, thus inducing $\Delta
m_F = \pm 1$ Raman transitions. Depending on the sign of the
pump-probe detuning $\delta$, the population imbalance induces
gain or absorption. Each pair of neighboring sublevels contributes
with a relative weight depending on the population inversion. In
practice however, the levels are not separated enough to be well
resolved, and only two structures (with opposite signs) are
visible, one corresponding to amplification for
$\delta=-\delta_\mathrm{R}$ and one to absorption for
$\delta=\delta_\mathrm{R}$. Note that this situation corresponds
to a red detuning for the pump ($\Delta < 0$) and that the signs
are inverted for blue-detuning ($\Delta > 0$). As
$\delta_\mathrm{R}$ comes from a differential light-shift (because
of different Clebsch-Gordan coefficients), it is usually on the
order of $\Gamma/10$, whereas $\Delta$ is a few $\Gamma$. The
width $\gamma$ of the resonances is related to the elastic
scattering rate, also much smaller than $\Gamma$
\cite{Grison:1991}. Far from the main atomic absorption resonance,
the Raman resonance is thus a narrow spectral feature, as in Fig.
\ref{fig.Raman}(c), such that we can fit it independently of the
main absorption line [Fig. \ref{fig.fit}(a)]. Therefore, we scan
the frequency of the probe beam around $\delta=0$ only, which
reduces the interaction time with the pump, thus suppressing
radiation pressure and subsequent unwanted Doppler shift. Note
that adding a second counterpropagating pump beam is not a
suitable solution, as in this situation, other mechanisms may
occur (recoil-induce resonances, four-wave mixing, atom
localization in potential wells)
\cite{Brzozowski:2005,Grynberg:2001,Chen:2001}, which would
complicate the analysis. We use the polarizability
$\tilde{\alpha}_\mathrm{R}(\delta, \Delta, \Omega)$ to describe
the Raman structure, with
\begin{equation} \label{eq.alpha_R_im}
\mathrm{Im}(\tilde{\alpha}_\mathrm{R}) =
\frac{A_1}{(\delta-\delta_\mathrm{R})^{2}+\gamma^{2}/4} -
\frac{A_2}{(\delta+\delta_\mathrm{R})^{2}+\gamma^{2}/4} \; .
\end{equation}
This function is particularly convenient as the Kramers-Kronig
transformation of a Lorentzian profile is well known. We thus
avoid any numerical integration.

Our experimental procedure is the following. We scan the probe
frequency from $\delta = -\Gamma$ to $\delta = \Gamma$ during
100~$\mu$s and record one Raman transmission spectrum. During the
same cycle, we perform two larger scans without pumping, one
before the pump-probe spectroscopy and one after, in order to
record the main absorption line (as in Fig. \ref{fig.Raman}(c)),
from which we extract the on-resonance optical thickness $b_0$.
The second measurement allows us to take into account the losses
induced by the pump radiation pressure. The corresponding
uncertainty on $b_0$ induces, at the end, a $\pm 10 \%$
uncertainty on the critical optical thickness. Then, we fit by
\begin{equation}\label{eq.fit} T(\delta)=\exp\left[ -b_0 \times
\left( \mathrm{Im}[ \tilde{\alpha}_\mathrm{R}(\delta)] + m\delta+p
\right) \right] \;,
\end{equation} where $A_1$, $A_2$, $\delta_\mathrm{R}$ and
$\gamma$ are the adjustable parameters of the Raman structure
described by $\tilde{\alpha}_\mathrm{R}$, and the adjustable line
parameterized by $m,p$ is used to fit the background of the
transmission spectrum. This expression is not rigorous and one
could instead search for the complete expression of the atomic
response. However, it is difficult to take into account the
complete real system, including the Zeeman degeneracy and
polarization effects. Using Eq. (\ref{eq.fit}) allows us to
efficiently measure the Raman parameters. As shown in Fig.
\ref{fig.fit}(a), the fit is very satisfactory. For most
parameters, the width $\gamma$ of the Lorentzians is larger than
their shift $\delta_\mathrm{R}$, so that the two Lorentzians are
not separated and the Raman structure looks like a dispersion
profile. Similarly, the corresponding scattering cross-section
looks like one bell-shaped curve. The obtained widths $\gamma$ are
consistent with the pump elastic scattering rate. The ratio
between the gain amplitude $A_1$ and the absorption amplitude
$A_2$ is approximately constant, as expected, since it depends
only on the Clebsch-Gordan coefficients.

Then, the Lorentzian shape of the Raman contribution to the atomic
polarizability is analytically transformed through Kramers-Kronig
relations to get
\begin{equation} \label{eq.alpha_R_re}
\mathrm{Re}(\tilde{\alpha}_\mathrm{R}) = A_1 \times \frac{
-2(\delta-\delta_\mathrm{R})/\gamma}{(\delta-\delta_\mathrm{R})^{2}+\gamma^{2}/4}
- A_2\times \frac{-2(\delta+\delta_\mathrm{R})/\gamma
}{(\delta+\delta_\mathrm{R})^{2}+\gamma^{2}/4} \; .
\end{equation}
The atomic polarizability $\tilde{\alpha}_\mathrm{R}$ is thus
fully determined.

However, the measurement is valid for the special polarization
configuration that we have used, whereas for a random laser, the
polarization is \textit{a priori} random. To get a realistic
estimation of the random laser threshold, we thus have to make an
average over the polarization. We have checked experimentally that
the coefficients $A_1$, $A_2$ have a $\sin^2(\theta)$ dependence
with the relative angle $\theta$ between the pump and the probe
linear polarizations. As we have performed all the measurements in
the optimum case (with the probe polarization perpendicular to the
pump one), it is appropriate to multiply the measured values of
$\mathrm{Im}(\tilde{\alpha}_\mathrm{R})$ by $1/2$ and
$|\tilde{\alpha}_\mathrm{R}|^2$ by $3/8$ (average of
$\sin^4(\theta)$). The cross-sections used to determine the random
laser threshold are thus
\begin{eqnarray}
\sigma_\mathrm{g}/\sigma_0  = &
\frac{3}{8}|\tilde{\alpha}_\mathrm{R}|^2 &-
\frac{1}{2}\mathrm{Im}(\tilde{\alpha}_\mathrm{R}) \, , \label{eq.sigma_g} \\
\sigma_\mathrm{sc}/ \sigma_0  =&
\frac{3}{8}|\tilde{\alpha}_\mathrm{R}|^2 &
\; , \label{eq.sigma_sc}
\end{eqnarray}
where $\tilde{\alpha}_\mathrm{R}$ is experimentally determined as
described above [Eqs. (\ref{eq.alpha_R_im}-\ref{eq.alpha_R_re})
and Fig. \ref{fig.fit}(a)]. An exemple of computed cross-sections
is shown in Fig. \ref{fig.fit}(b).

Then, the critical optical thickness is easily computed from
\begin{equation}\label{eq.b0crbis}
\eta b_{0\mathrm{cr}} =
\frac{2\pi\sigma_0}{\sqrt{3\sigma_\mathrm{sc}\sigma_\mathrm{g}}}
\; ,
\end{equation}
where the correcting $\eta$ factor (coming from the extrapolation
length) writes $\eta=
\left(b_{0\mathrm{cr}}\sigma_\mathrm{sc}+4\zeta\sigma_0\right)/\left(b_{0\mathrm{cr}}\sigma_\mathrm{sc}+2\zeta\sigma_0\right)$
and yields a second-order equation in $b_{0\mathrm{cr}}$. The
solution, plotted as a function of $\delta$, is reported on Fig.
\ref{fig.fit}(c). As expected, the minimum is located near the
maximum of the gain cross-section, \textit{i.e.} for $\delta
\simeq -\delta_\mathrm{R}$.

\section{Results and discussion}

\begin{figure}[t]
\centering
\includegraphics{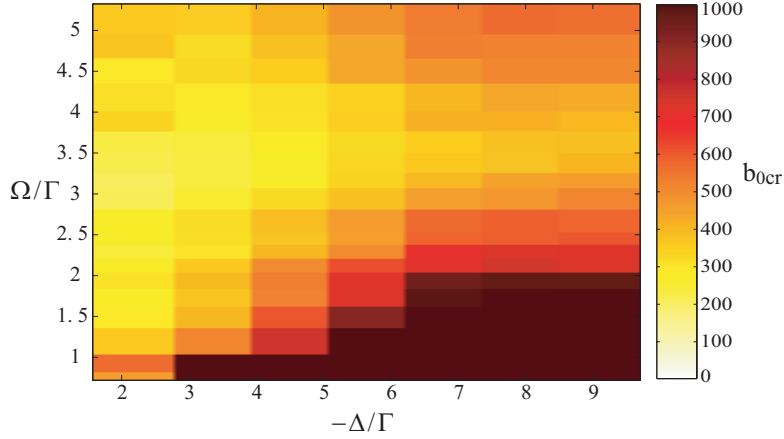}
\caption{Critical optical thickness $b_{0\mathrm{cr}}$ as a
function of the pumping parameters $\Delta$ (atom-pump detuning)
and $\Omega$ (Rabi frequency of the atom-pump coupling). The
minimum is around $b_{0\mathrm{cr}} \sim 210-230$, for $\Delta
\sim 2 \Gamma$ and $\Omega \sim 2-3 \Gamma$.}\label{fig.b0cr}
\end{figure}

We repeat the above procedure for each couple of pumping
parameters $\{\Delta,\Omega\}$. The Rabi frequency of the
atom-pump interaction has been calibrated by monitoring the light
shift of the main absorption line \cite{Cohen:PetitLivreRouge-EN}
as a function of the pump intensity, as it can be seen in Fig.
\ref{fig.Raman}(c). We studied only the $\Delta<0$ part, as Raman
gain is independent on the sign of $\Delta$ \cite{Chen:2001}.
Moreover, we have been restricted to $|\Delta| \geq 2$ because too
much radiation pressure destroys the MOT for $|\Delta|<2$. As the
random laser will automatically start with the first frequency
above threshold, we report in Fig. \ref{fig.b0cr} the critical
optical thickness defined as
\begin{equation}
b_{0\mathrm{cr}}(\Delta,\Omega) = \min_\delta
[b_{0\mathrm{cr}}(\delta,\Delta,\Omega)].
\end{equation}
The minimum is around $b_{0\mathrm{cr}} \sim 210-230$, obtained for $\Delta
\sim 2 \Gamma$ and $\Omega \sim 2-3 \Gamma$.

Once the critical optical thickness is computed, the
self-consistency of our model has to be checked on two points.
Firstly, the diffusive approach leading to Eq. (\ref{eq.letokhov})
requires in principle that the ratio $L/\ell_\mathrm{sc} =
b_{0\mathrm{cr}}\times\sigma_\mathrm{sc}/\sigma_0$ is
substantially larger than one to be justified. Nevertheless, it
has been shown recently that this condition has not to be strictly
respected, as the diffusive approach gives quite accurate results
down to $L/\ell_\mathrm{sc} \sim 1$ \cite{Froufe:2009}. This is
approximately the value obtained for the optimum parameters. Note
also that the correction due to the extrapolation length ($\eta$
factor) is not negligible, as for $L/\ell_\mathrm{sc} \sim 1$,
$\eta \sim 1.6$.

Secondly, we have so far only considered the Raman resonance,
neglecting the influence of the main atomic transition at
$\omega_0$, which is valid for very large detunings $\Delta \gg
\Gamma$. However, since the optimum threshold is obtained for
small detuning, this is not justified. The corresponding
one-photon transition has no gain around $\delta = 0$ and then
only adds scattering. This scattering can be decomposed into
elastic and inelastic contributions. The elastic contribution
will lower the random laser threshold, whereas the inelastic
contribution, which shifts the frequency out of the Raman gain
curve, will yield an increase of the lasing threshold. Let us
examine the effect of the supplementary elastic scattering. It
can be evaluated by \begin{equation} \label{eq.sigma2level}
\sigma_\mathrm{el} = \frac{\sigma_0}{1+4\Delta^2/\Gamma^2} \times
\frac{1}{(1+s)^2} \times \mathcal{C}\; . \end{equation} The first
term is the total scattering cross-section of a two-level atom,
taking into account the detuning. The second factor, where
$s=2\Omega^2/(\Gamma^2+4\Delta^2)$ is the pump saturation
parameter, describes the reduced scattering cross-section,
keeping only the elastic part \cite{Cohen:PetitLivreRouge-EN}. As
a change of Zeeman sublevel is possible during a scattering
event, an additional weighting factor, estimated as $\mathcal{C}
\sim 0.5$ \cite{Gao:1994}, is necessary to select true elastic
scattering. Adding $\sigma_\mathrm{el}$ to $\sigma_\mathrm{sc}$
[Eq. (\ref{eq.sigma_sc})] lowers the critical optical thickness
to $b_{0\mathrm{cr}} \sim 120-130$, with approximately the same
optimum pumping parameters. This is however an optimistic
evaluation, as inelastic scattering has not been taken into
account. On the contrary, a conservative evaluation can be
obtained by considering inelastic scattering as pure losses,
\textit{i.e.} as a negative contribution to the gain
cross-section. This is pessimistic because those photons may not
be definitively lost, as further inelastic scattering can shift
their frequency back on the gain curve. As previously, the
inelastic scattering cross-section can be evaluated by
\begin{equation} \label{eq.sigma_inel} \sigma_\mathrm{inel} =
\frac{\sigma_0}{1+4\Delta^2/\Gamma^2} \times
\left[\frac{1}{(1+s)^2} \times (1-\mathcal{C}) +
\frac{s}{(1+s)^2}\right]\; . \end{equation} The first term in the
squared bracket is associated with Raman inelastic scattering
whereas the second term is due to incoherent scattering of the
two-level atom \cite{Mollow:1969}. Subtracting
$\sigma_\mathrm{inel}$ to the gain cross-section of Eq.
(\ref{eq.sigma_g}) increases now the critical optical thickness to
$b_{0\mathrm{cr}} \sim 215 - 230$. The optimum parameters are then
located near $\Omega \sim 3-4 \Gamma$ and $\Delta \sim 3-4 \Gamma$. Except for
small $\Delta$, where inelastic scattering is dramatic, the result
is not very different from the one presented on Fig.
\ref{fig.b0cr}. Especially near the optimum parameters ($\Omega =
\Delta \sim 3-4 \Gamma$), the optimistic evaluation leads to
$b_{0\mathrm{cr}} \sim 165 - 180$, which is not very different
from the pessimistic result ($b_{0\mathrm{cr}} \sim 215 - 230$).
Therefore, we conclude that the value $b_{0\mathrm{cr}} \sim 200$
gives the correct order of magnitude. Such a high optical
thickness is achievable, for instance by using compression
techniques of magneto-optical traps
\cite{Ketterle:1993,DePue:2000}. The corresponding ratio
$L/\ell_\mathrm{sc}$ is on the order of 2.

Random lasing occurs at a detuning from the pump $|\delta| \sim
\delta_\mathrm{R}$, typically smaller than 1~MHz. This makes the
detection of such a random laser very challenging, as the
corresponding fluorescence cannot easily be separated from the
pump-induced fluorescence. Nevertheless, the narrow Raman
structure could be revealed by a beat note experiment, as in
\cite{Westbrook:1990}, or alternatively by the intensity
correlations in the fluorescence, measured either by a homodyne
technique \cite{Jurczak:1995} or with a time correlator
\cite{Bali:1996}. In this last experiment, a contribution from
Raman scattering has been measured, consistent with the
theoretical predictions of \cite{Gao:1994}. It seems reasonable to
expect this signal to have different behaviors below and above
threshold, but this remains to be checked by further theoretical
studies.


Finally, let us mention that our model contains several
limitations, so that the numbers should be considered as
first-order estimates. Our description of Raman gain is quite
simplified in order to have an efficient data analysis procedure,
leading to quasi-analytical results. Precise modelling of the
complete atomic response is indeed not the goal of this article.
On the contrary, Raman gain is used as a convenient example to
illustrate the method, which is general and could be used with any
gain mechanism, by numerically computing the real part of the
atomic polarizability from the experimental transmission spectrum,
via Kramers-Kronig relations.

Our hypothesis of homogeneous atomic density and monochromatic and
homogeneous pumping could also be discussed \cite{Noginov:2006},
especially when high optical thicknesses are involved, as the pump
attenuation may become important. These effects could be taken
into account in numerical simulations of light transport in active
and disordered medium, but have to be neglected to allow the
analytical resolution of the diffusion equation leading to Eq.
(\ref{eq.letokhov}) \cite{Letokhov:1968,Cao:2003}. Note however
that the on-resonance optical thickness is not the relevant
parameter for the pumping field, since the pump is detuned and is
saturating. Moreover, diffused pump light, which penetrates into
the sample much deeper than the coherent transmission, has also to
be taken into account.


\section{Conclusion}

We have presented a method to experimentally determine the
threshold of random lasing in a cloud of cold atoms. In this
specific system, the threshold is related only to the complex
atomic polarizability, which can be fully characterized by
spectroscopic measurements. We applied this idea with Raman gain
between light-shifted Zeeman sublevels of rubidium atoms. From our
measurements, we estimate the critical optical thickness to be on
the order of 200, which is achievable with current cold-atom
experiments.

The obtained critical optical thickness is lower than the one
obtained with Mollow gain \cite{Froufe:2009}. This is in agreement
with the intuition that more complex gain mechanisms offer more
degrees of freedom, which is of course necessary to optimize
several quantities (scattering and gain) at the same time. We are then confident that even lower thresholds can be
obtained with other, more complex gain mechanisms, for example
non-linear parametric gain induced by non-degenerate four-wave
mixing. This may be the subject of our future investigations.

\ \\

\noindent \textbf{Acknowledgments}\\

\noindent We thank F. Michaud and R. Carminati for fruitful
discussions. We acknowledge financial support from the program
ANR-06-BLAN-0096, funding for N.M. by DGA and for D.B. by
INTERCAN.

\end{document}